\documentclass[pre,aps,floatfix,notitlepage]{revtex4-1}

\usepackage[dvips]{graphicx}
\usepackage{amssymb,amsfonts,amsmath,calc,ifsym,bbding,wasysym}
\usepackage[usenames]{color}
\usepackage{bm}
\usepackage[normalem]{ulem}
\usepackage{xr}
\usepackage{float}
\usepackage{grfext}
\usepackage{braket}
\PrependGraphicsExtensions*{.pdf,.PDF}  

\definecolor{matt}{rgb}{0.27,0,0.51}

\def\ld{\lambda_\text{D}}
\def\Csalt{C_\text{salt}}
\def\I{I}
\def\ess{\varepsilon_\text{ss}}
\def\gss{g_\text{ss}}
\newcommand{\kd}{K_\mathrm{d}}
\newcommand{\sbb}{s_\mathrm{b}}
\newcommand{\scc}{s_\mathrm{c}}

\newcommand{\kt}{k_\mathrm{B}T}

\def\eunit{E_\text{u}}

\def\dunit{D_\text{u}}

\def\qp{q_\text{p}}

\def\rcut{r_\text{cut}}
\newcommand{\sub}[2]{ _{\mathrm{#1}#2}}
\newcommand{\rsub}[1]{_\mathrm{#1}}

\newcommand{\LJ}[1]{ \mathcal{L}_{#1} }
\newcommand{\Morse}[1]{ \mathcal{M}_{#1} }

\newcommand{\tobs}{t_\text{f}}
\newcommand{\nads}{n_\text{ad}}
\newcommand{\nfree}{n_\text{free}}
\newcommand{\nfreebar}{\bar{n}_\text{free}}
\newcommand{\ceq}{c_\text{eq}}
\newcommand{\cstar}{c_\text{eq}^*}
\newcommand{\Istar}{C_\text{salt}^*}

\newcommand{\Rg}{R_\text{g}}
\newcommand{\Rh}{R_\text{H}}

\begin{document}

\renewcommand{\figurename}{Abstract Figure}

\title{Pathways for virus assembly around nucleic acids}
\author{Jason D Perlmutter}
\author{Matthew R Perkett}
\author{Michael F Hagan}
\email{hagan@brandeis.edu}
\affiliation{Martin Fisher School of Physics, Brandeis University,
  Waltham, MA, USA.}
\maketitle


\section{Abstract}
Understanding the pathways by which viral capsid proteins assemble around their genomes could identify key intermediates as potential drug targets. In this work we use computer simulations to characterize assembly over a wide range of capsid protein-protein interaction strengths and solution ionic strengths. We find that assembly pathways can be categorized into two classes, in which intermediates are either predominantly ordered or disordered. Our results suggest that estimating the protein-protein and the protein-genome binding affinities may be sufficient to predict which pathway occurs. Furthermore, the calculated phase diagrams suggest that knowledge of the dominant assembly pathway and its relationship to control parameters could identify optimal strategies to thwart or redirect assembly to block infection. Finally, analysis of simulation trajectories suggests that the two classes of assembly pathways can be distinguished in single molecule fluorescence correlation spectroscopy or bulk time resolved small angle x-ray scattering experiments.



\setcounter{figure}{0}
\renewcommand{\figurename}{Figure}

\pagebreak
\section{Introduction}

In many virus families, the spontaneous assembly of a protein shell (capsid) around the viral genome is an essential step in the viral life cycle \cite{Hagan2014}. These families include most viruses with single stranded RNA (ssRNA) genomes, as well as the Hepadnaviridae (e.g. hepatitis B virus, HBV). Understanding the mechanisms which underlie this cooperative assembly process could facilitate efforts to develop anti-viral drugs that block or derail the formation of infectious particles (for reviews see \cite{Zlotnick2011, Prevelige2011}) and promote efforts to reengineer them for biomedical delivery. In this article, we explore how the interactions between the molecular components determine the mechanism of assembly, and how these interactions can be altered by changing solution conditions or mutagenesis to modulate assembly pathways.

The most detailed knowledge of capsid-RNA interactions comes from structural analysis of assembled viral particles. Atomic resolution structures of capsids assembled around ssRNA have been obtained by x-ray crystallography and/or cryo-electron microscopy (cryo-EM) (e.g. \cite{Fox1994,Valegard1997,Johnson2004,Tihova2004,Krol1999,Stockley2007,Toropova2008,Lucas2002,Valegard1994,Worm1998,Grahn2001,Valegard1997,Helgstrand2002,Schneemann2006}).
The packaged NAs are less ordered than their protein containers and thus have been more difficult to characterize. However cryo-EM experiments have identified that the nucleotide densities are nonuniform, with a peak near the inner capsid surface and relatively low densities in the interior\cite{Tihova2004,Zlotnick1997,Conway1999}. While atomistic detail has not been possible in these experiments, all-atom models have been derived from equilibrium simulations \cite{Freddolino2006,Devkota2009,Zeng2012}. In some cases, striking image reconstructions reveal that the packaged RNA adopts the symmetry of the overlying capsid (e.g. \cite{Tihova2004,Toropova2008,Schneemann2006,Bakker2012,Ford2013}). While it has been proposed that this order arises as a function of the assembly mechanism for several viruses \cite{Larson2001, Dykeman2011, Dykeman2013}, computational analysis of polyelectrolyte configurations inside capsids also indicate that capsid-polymer interactions can generically drive spatial organization of the packaged polymer \cite{Siber2008,Hu2008c,Devkota2009,Forrey2009, Harvey2009,Belyi2006, Angelescu2006,Zhang2004a,Lee2008,Angelescu2008,Jiang2009, Elrad2010,Perlmutter2013}. Theoretical works have also characterized the relationship between the NA charge and structure and the length which is optimal for packaging \cite{Perlmutter2013,Schoot2013,Hu2008a,Belyi2006, Schoot2005, Angelescu2006, Siber2008,Ting2011, Ni2012,Siber2010,Siber2012}.

In addition to this structural data on assembled capsids, an extensive combination of mass spectrometry, assembly kinetics experiments, constraints from assembled capsid structures, and mathematical modeling has delineated assembly pathways for several viruses, with a particular focus on the role of interactions between CPs and specific RNA sequences called packaging signals. Recent single molecule fluorescence correlation spectroscopy (smFCS) experiments indicate that, for these viruses, assembly around the viral genome is more robust and proceeds by a different mechanism as compared to around heterologous RNA \cite{Borodavka2012}. Yet, in other cases capsid proteins show no preference for genomic RNA over heterologous RNA (e.g. HBV \cite{Porterfield2010}), and cowpea chlorotic mottle virus (CCMV) proteins preferentially encapsidate heterologous RNA (from BMV) over the genomic CCMV RNA with equivalent length \cite{Comas-Garcia2012}. Furthermore, experimental model systems in which capsid proteins assemble into icosahedral capsids around synthetic polyelectrolytes or other polyanions \cite{Hiebert1968,Bancroft1969,Dixit2006,Loo2006,Goicochea2007,Huang2007,Loo2007,Sikkema2007,Sun2007,Hu2008,Comellas-Aragones2009,Crist2009,Chang2008}) demonstrate that specific RNA sequences are not required for capsid formation or cargo packaging.
Thus, a complete picture of capsid assembly mechanisms requires understanding how assembly pathways depend on those features which are generic to polyelectrolytes, as well as those which are specific to viral RNAs.

In previous work on assembly around a simple model for a polymer, Elrad and Hagan \cite{Elrad2010} proposed that mechanisms for assembly around a cargo (i.e. RNA, polymer, or nanoparticle) can be classified on the basis of two extreme limits.  In the first (originally proposed by McPherson \cite{McPherson2005} and then Refs \cite{Hagan2008,Devkota2009}), strong protein-cargo interactions drive proteins to adsorb \textit{`en masse'} onto the cargo in a disordered manner, meaning there are few protein-protein interactions. Once enough subunits are bound, subunits undergo cooperative rearrangements (potentially including dissociation of excess subunits) to form an ordered capsid. This mechanism has been observed in recent simulations \cite{Mahalik2012, Perlmutter2013, Elrad2010, Hagan2008, Perkett2014}. In the second limit, where protein-protein interactions dominate, a small partial capsid nucleates on the cargo, followed by a growth phase in which individual proteins or small oligomers sequentially add to the growing capsid. This class of pathways resembles the nucleation-and-growth mechanism by which empty capsids assemble \cite{Endres2002}, except that the polymer plays an active role by stabilizing protein-protein interactions and by enhancing the flux of proteins to the assembling capsid \cite{Hu2007,Kivenson2010,Elrad2010}.

It is difficult to determine assembly mechanisms directly from experiments, due to the small size ($\lesssim10$ of nm) and transience ($\sim$ms) of most intermediates. Observations \textit{in vitro} suggest that both mechanisms may be viable. Kler et al. \cite{Kler2012} used time-resolved X-ray scattering (trSAXS) to monitor SV40 capsid proteins assembling around ssRNA. The scattering profiles at all time points during assembly could be decomposed into unassembled components (RNA + protein subunits) and complete capsid; the absence of any signal corresponding to a large disordered intermediate suggests this assembly follows the nucleation-and-growth (ordered) assembly mechanism \cite{Kler2012}.
Other observations suggest that viruses can assemble through the en masse mechanism. Refs. \cite{Garmann2013,Cadena-Nava2012} found that \textit{in vitro} assembly CCMV assembly was most productive when performed in two steps: (1) at low salt (strong protein-RNA interactions) and neutral pH (weak protein-protein interactions) the proteins undergo extensive adsorption onto RNA, then (2) pH is reduced to activate binding of protein-protein binding \cite{Garmann2013}. Similarly, a recent observation of capsid protein assembly around charge-functionalized nanoparticles found that assembly initially proceeded through nonspecific aggregation of proteins and nanoparticles, followed by the gradual extrusion of nanoparticles within completed capsids \cite{Malyutin2013}. These experiments used viral proteins with relatively weak protein-protein interactions (CCMV and BMV) \cite{Zlotnick2013} and moderate salt concentrations ($100-150$mM). The Kler et al. \cite{Kler2012,Kler2013} experiments considered SV40 proteins, which have strong protein-protein interactions \cite{Zlotnick2013}, and high salt ($250$mM) Together, these \textit{in vitro} experiments suggest that productive assembly could proceed by either the en masse or nucleation-and-growth mechanism.  

In this work, we use dynamical simulations to investigate the extent to which the assembly mechanism can be controlled by tuning solution ionic strength and protein-protein attractions. We extend a model that was recently used to calculate the thermostability and assembly yields of viral particles as a function of protein charge and nucleic acid (NA) length and structure. Those previous simulations found quantitative agreement between predicted NA lengths that optimize capsid thermostability and viral genome length for seven viruses \cite{Perlmutter2013}. Here, we perform extensive new simulations of assembly, in which protein-protein interactions, the sequence of charges in capsid protein-NA binding domains, and the solution ionic strength are varied. We find that, by varying these control parameters, the assembly mechanism can be systematically varied between the two extreme limits described above. Our results suggest that knowledge of protein-protein and protein-NA binding affinities may be sufficient to predict which assembly mechanism will occur, and we estimate relative protein-NA binding interactions for three viruses (based on non-specific electrostatic interactions).  These findings suggest that assembly mechanisms can be rationally designed through choice of solution conditions and mutagenesis of capsid protein-protein interfaces and protein-NA binding domains.
 Finally, by calculating hydrodynamic radii and SAXS profiles associated with assembly intermediates, we show that assembly mechanisms can be distinguished by experimental techniques recently applied to virus assembly, respectively  smFCS \cite{Borodavka2012}) and trSAXS \cite{Kler2012}.  While the NA is represented by a linear polyelectrolyte in most of the simulations, we obtain qualitatively similar results when considering a model for base-paired NAs developed in Ref. \cite{Perlmutter2013}.


\section{Results}
\begin{figure}
\centering{\includegraphics[width=0.5\columnwidth]{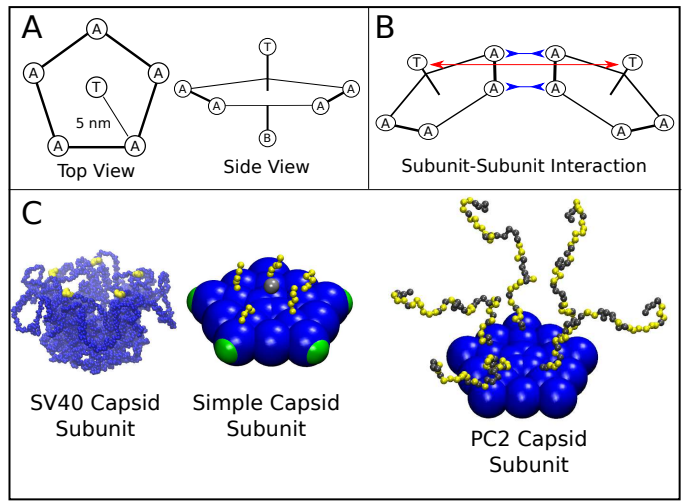}}
\caption{{\bf (A),(B)} Model schematic for (A) a single subunit, and (B) two interacting subunits, showing positions of the attractor (`A'), Top (`T'), and Bottom (`B') pseudoatoms, which are defined in the Model section and in the Methods. {\bf (C)} (left) The pentameric SV40 capsid protein subunit, which motivates our model. The globular portions of proteins are shown in blue and the beginning of the NA binding motifs (ARMs) in yellow, though much of the ARMs are not resolved in the crystal structure \cite{Stehle1996}. Space-filling model of the generic subunit model (middle) and a pentamer from the PC2 model (right). Beads are colored as follows: blue=excluders, green=attractors, yellow=positive ARM bead, gray=neutral ARM bead, red=polyelectrolyte.
\label{schematic}
}
\end{figure}

To study how capsid assembly around a polyelectrolyte depends on the strength of protein subunit-subunit and subunit-polyelectrolyte interactions, we performed Brownian dynamics simulations with a recently developed model \cite{Perlmutter2013} (Figure~\ref{schematic}).
This model was motivated by the observation \cite{Kler2012, Kler2013} that purified simian virus 40 (SV40) capsid proteins assemble around ssRNA molecules to form capsids composed of 12 homopentamer subunits. The capsid is modeled as a dodecahedron composed of 12 pentagonal subunits (each of which represents a rapidly forming and stable pentameric intermediate, which then more slowly assembles into the complete capsid, as is the case for SV40 \textit{in vitro} around ssRNA \cite{Li2002, Kler2012, Kler2013}). Subunits are attracted to each other via attractive pseudoatoms at the vertices (type `A') and driven toward a preferred subunit-subunit angle by repulsive `Top' pseudoatoms (type `T') and `Bottom' pseudoatoms (type `B') (see Fig.~\ref{schematic} and the Methods section). These attractions represent the interactions between capsid protein subunits that arise from hydrophobic, van der Waals, and electrostatic interactions as well as hydrogen bonding \cite{Hagan2014}). The magnitude of these interactions varies between viral species \cite{Zlotnick2013} and can be experimentally tuned by pH and salt concentration \cite{Ceres2002,Kegel2004,Hagan2014}. Here, the attraction strength is controlled by the model parameter $\ess$. In order to relate this potential energy to the free energy of dimerization, we have run a separate series of calculations, where in the absence of cargo, we find that the free energy of subunit of dimerization is $\gss{=}5.0-1.5*\ess$ (see SI section~\ref{Uaa}). Throughout this article energies are given in units of the thermal energy, $\kt$.

Capsid assembly around nucleic acids and other polyelectrolytes is primarily driven by electrostatic interactions between negative charges on the encapsulated polyelectrolyte and positive charges on the inner surfaces of capsid proteins \cite{Hagan2014}. We consider a linear bead-spring polyelectrolyte, with a charge of $–e$ per bead and a persistence length comparable to that of ssRNA in the absence of base pairing. Positive charges on capsid proteins are located in flexible polymers affixed to the inner surface of the model subunit, which represent the highly charged, flexible terminal tails known as arginine rich motifs that are typical of positive-sense ssRNA virus capsid proteins (e.g., \cite{Schneemann2006}).
Electrostatics are modeled using Debye-Huckel interactions, where the Debye screening length ($\ld$) is determined by the ionic strength $\I$, equivalent to the concentration $\Csalt$ of a monovalent salt, as $\ld \approx 0.3/\I^{1/2}$ with $\ld$ in nm and $\I$ in molar units. Perlmutter et al. \cite{Perlmutter2013} showed that Debye-Huckel interactions compare well to simulations with explicit counterions for the parameter values under consideration; furthermore, inclusion of multivalent ions at experimentally relevant concentrations had only a small effect on properties such as the optimal genome length. The effect of base-pairing on the optimal length was also considered in that reference \cite{Perlmutter2013}.

\subsection{Kinetic phase diagram}
\begin{figure}
\centering{\includegraphics[width=0.9\columnwidth]{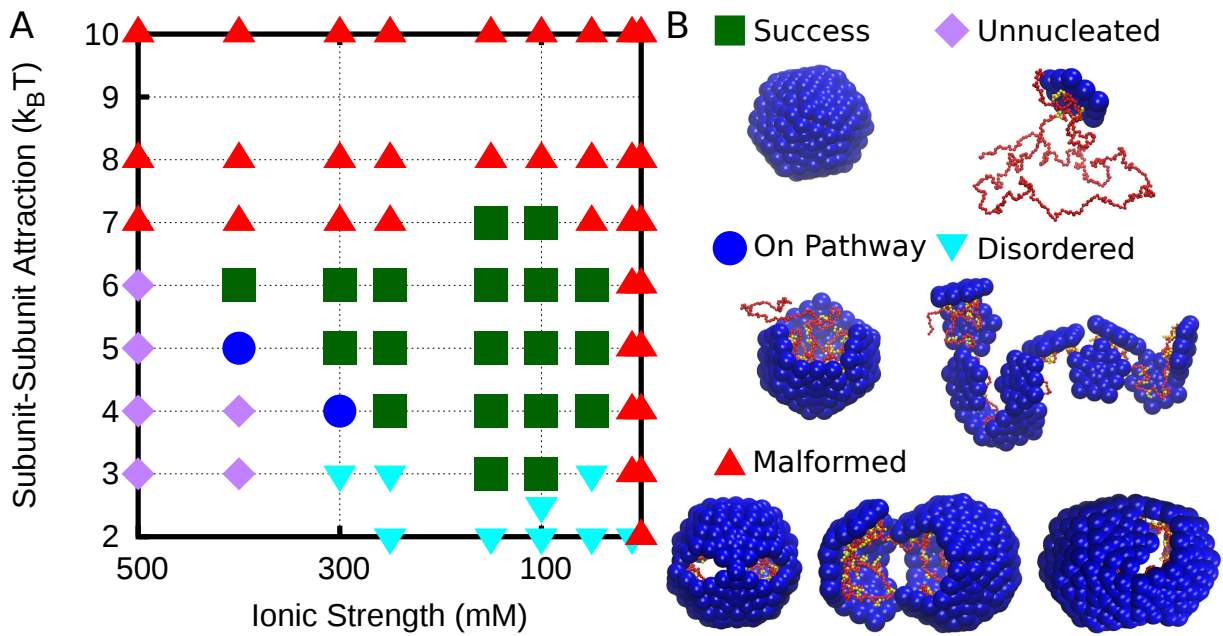}}
\caption{{\bf (A)} Kinetic phase diagram showing the most prevalent final product at the conclusion of assembly simulations ($\tobs {=}2\times 10^8$ time steps). {\bf (B)} Snapshots illustrating categories.
\label{outcome}
}
\end{figure}

\begin{figure}
\centering{\includegraphics[width=0.7\columnwidth]{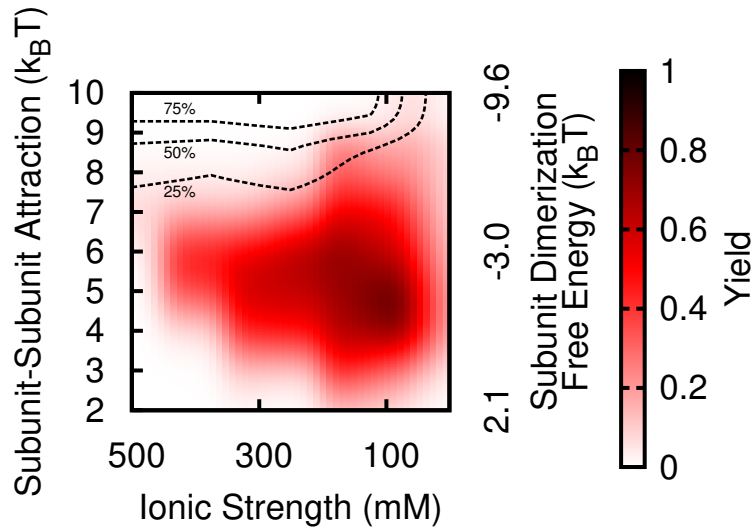}}
\caption{Kinetic phase diagram showing fraction of simulations which result in successful assembly of a complete capsid. The black isosurface lines show the fraction of subunits which are not adsorbed to the polymer and show any partial assembly, i.e. above the $75\%$ line, $\geq 75\%$ of the subunits not bound to the polymer are clustered.
\label{yield}
}
\end{figure}

 We first consider the predominant assembly products (Fig.~\ref{outcome}) and the packaging efficiency (Fig.~\ref{yield}) as a function of Debye length $\ld$ and subunit-subunit interaction strength $\ess$. The packaging efficiency is defined as the fraction of trajectories in which the polyelectrolyte is completely encapsulated by a well-formed capsid, which contains 12 subunits each of which interact with 5 neighbors. We refer to this as a kinetic phase diagram \cite{Hagan2006, Hagan2014} because we characterize products at a finite observation time of $\tobs{=}2\times10^8$ time steps, which is long enough that assemblages do not vary significantly with time (except for under weak interactions, see below), but is not sufficient to equilibrate kinetic traps if there are large activation barriers \cite{Elrad2010, Hagan2014}. We see that for the range of simulated ionic strengths ($1 \le \Csalt \le 500$ mM or $10 \ge \ld \ge 0.4$ nm) assembly yields are highest for $\ess{=}5\kt$ and $\Csalt{=}100$ mM (the parameter values focused on in \cite{Perlmutter2013}), and that for moderate subunit-subunit interaction strengths ($4\le\ess\le6, \kt$) yield to remain high as the ionic strength is increased to about 300 mM ($\ld\approx0.6$ nm). For higher salt concentrations, yields are depressed by the appearance of long-lived on-pathway intermediates. As will be discussed further below, weakening the electrostatic interaction between the polymer and protein limits the ability of the polymer to promote assembly. Although we expect that these simulations would eventually result in complete capsids, the low yield at our finite measurement time reflects the fact that assembly is less efficient than for lower salt concentrations. At the highest ionic strength considered ($\Csalt{=}500$ mM), the most prevalent outcome is that no nucleation occurs. At lower salt concentrations ($\Csalt\ge10$ mM), rapid adsorption of a super-stroichiometric quantity of subunits results in malformed capsids.

At larger-than-optimal protein-protein interaction strengths ($\ess>6$) assembly yields are lower for two reasons. The first, and more frequent outcome, is long-lived malformed structures with strained interactions. This kinetic trap arises in a wide variety of assembly systems when interactions become strong in comparison to the thermal energy, because strained interactions are unable to anneal before additional subunits lock them in place \cite{Hagan2014, Hagan2011, Grant2011}. In our simulations, we found that these structures frequently result from an incorrect merger between two partial capsid intermediates; even when each individual interaction is too weak to lock-in the non-optimal structure, multiple erroneous interactions formed by two partial capsids are effectively irreversible on our timescale. The tendency for oligomer binding to lead to malformed structures was seen previously in the context of empty capsid assembly \cite{Hagan2006, Whitelam2009}. Here, the polymer helps to bring oligomers together, and thus this trap arises when nucleation on the polymer is faster than growth of a nucleus into a complete capsid. This trap resembles Geminivirus particles, which are composed of a union of two nearly complete capsids \cite{Bottcher2004}.

The second obstacle to polyelectrolyte encapsulation arises at the highest protein-protein interaction strengths studied ($\ess \geq 8\kt$), for which subunits not associated with the polyelectrolyte undergo spontaneous assembly. The resulting off-polyelectrolyte assembly depletes the pool of available monomers and small oligomers available for assembly on the polyelectrolyte, leading to a form of the monomer-starvation kinetic trap previously discussed for empty capsid assembly \cite{Hagan2014, Endres2002}. Triggering formation of empty capsids and thus preventing nucleic acid encapsidation by strengthening subunit-subunit interactions has been suggested as a mode of action for a putative antiviral drug for hepatitis B virus \cite{Katen2010, Katen2013}.

At smaller-than-optimal protein-protein interaction strengths ($\ess<4$) assembly is unsuccessful for two reasons, depending upon the ionic strength. At smaller values ($\Csalt \leq 300$ mM), electrostatic interactions are relatively strong, and many proteins adsorb onto to the polymer. However, because of the weak protein-protein interaction strength, these proteins do not form stable capsids, predominantly because nucleation is slow in comparison to $\tobs$ (the final observation time, $\tobs{=}2\times10^8$ time steps). In a minority of cases, a nucleus will form, but completion is prevented by the excess number of subunits adsorbed to the polyelectrolyte.
We refer to the resulting configurations as disordered, due to the lack of ordered binding between protein subunits. At larger ionic strengths ($\Csalt > 300$ mM) electrostatic interactions are relatively weak, and individual subunits rapidly desorb from the polyelectrolyte. In this regime, assembly requires a fluctuation in the number of adsorbed subunits which leads to nucleation of a partial capsid intermediate which has enough subunit-polyelectrolyte interactions to be stable against rapid desorption. The nucleation rate decreases exponentially with subunit-subunit interaction strength \cite{Hagan2014, Kivenson2010}, and thus most simulations at high salt and weak subunit-subunit interactions never undergo nucleation. (We categorize simulations with fewer than 3 subunits adsorbed to the polyelectrolyte and no subunit-subunit interactions (i.e. no progress towards assembly) as `unnucleated'.)

Importantly, we expect that trajectories in this region of parameter space will eventually undergo nucleation. Thus, as the finite assembly time $\tobs$ is increased the region of successful assembly will expand to lower values of $\ess$ and higher ionic strength, until eventually reaching values below which capsid assembly is thermodynamically unstable (see \cite{Hagan2006}, Fig. 7). To confirm this possibility, we used the Markov state model (MSM) approach described in Ref.~\cite{Perkett2014} to characterize assembly with $\ess{=}5,6$ and $\Csalt{=}500$ mM (see Figs.~\ref{snapsGrowth}C and \ref{twoState} (in the SI)). In contrast, the malformed capsids encountered under large parameter values typically will not resolve on any relevant timescales, since numerous strong interactions would need to be broken \cite{Hagan2014}. Thus, the boundaries of successful assembly at large $\ess$ values are insensitive to $\tobs$.


Several additional conclusions can be drawn from the variation of packaging efficiency (Fig.~\ref{yield}). Firstly, the yield of complete virus-like particles is relatively high for moderate subunit-subunit interaction strengths $\ess\in [4,6]$ across a range of ionic strengths ($\Csalt \in [100,300]$ mM). Even above ($\Csalt > 300$ mM) and below ($\Csalt{=}10$ mM) this range we observe moderate yields of complete particles. Secondly, as electrostatic interactions are weakened (moving to the left), the subunit-subunit interaction strength which optimizes the yield increases (i.e. from $\ess=5\kt$ at $\Csalt =100mM$ to $\ess=6\kt$ at $\Csalt: 150-400mM$ to $\ess=7\kt$ at $\Csalt=500mM$). This result suggests that one interaction type can compensate for the other within a limited range. However, though all successful capsids contain the same subunit geometry, the mechanism by which they form depends on the relative interaction strengths, as discussed next.

\subsection{Assembly mechanisms}
\begin{figure}
\centering{\includegraphics[width=0.9\columnwidth]{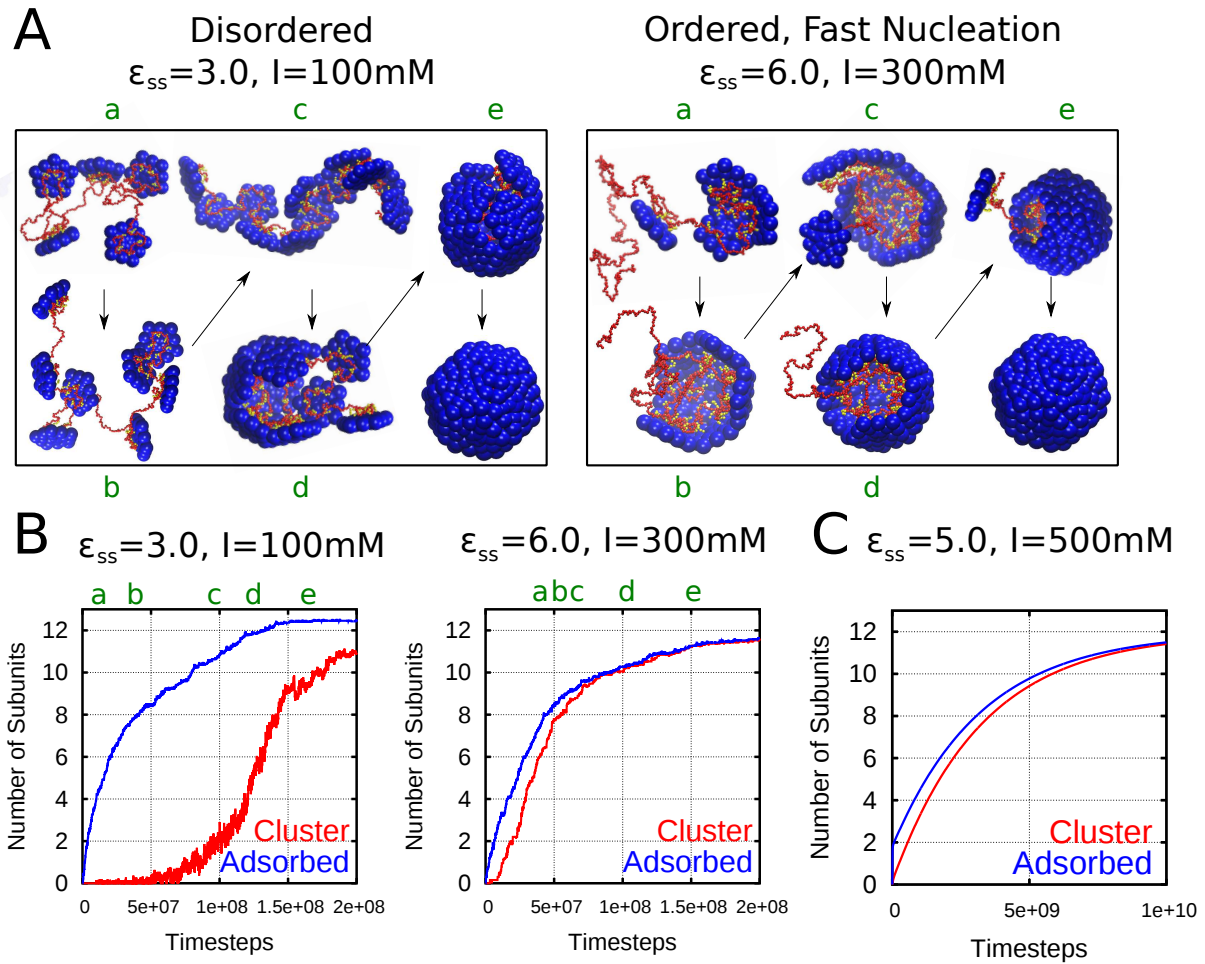}}
\caption{{\bf (A)} Snapshots from trajectories corresponding to the \textit{en masse} and nucleation-and-growth (ordered) assembly mechanisms, at indicated parameter values.   {\bf (B)} The number of subunits adsorbed ($\nads$) to the polyelectrolyte and the size of the largest cluster ($n$) are shown as a function of simulation time steps, averaged over all trajectories at the two sets of parameter values shown in {\bf (A)}.  The labels `a' through `e' connect each structure pictured in {\bf (A)} to its corresponding value of $\nads$ in {\bf(B)}.  {\bf (C)} Average values of $\nads$ and $n$ are shown as a function of time step for parameters that lead to the nucleation-and-growth mechanism with a larger nucleation barrier than in {\bf (B)}.
\label{snapsGrowth}
}
\end{figure}

 As noted previously \cite{Hagan2008, Elrad2010, Hagan2014}, pathways for assembly around a central core such as a polyelectrolyte can be roughly separated into two classes. In the first class (Fig.~\ref{snapsGrowth}a), which we refer to as the `\textit{en masse}' mechanism, subunits first adsorb onto the polyelectrolyte in a disordered manner, followed by cooperative rearrangements to form an ordered capsid. In the second class (Fig.~\ref{snapsGrowth}b), referred to as the nucleation-and-growth mechanism, a small partial capsid nucleates on the polyelectrolyte followed by the sequential, reversible addition of subunits or small oligomers until assembly completes. In contrast to the earlier models which considered a qualitative subunit-polyelectrolyte interaction, we study here how assembly pathways depend on the ionic strength.

To quantify the degree of order along assembly pathways, we record the total number of subunits adsorbed to the polyelectrolyte $\nads$ and the number of subunits in the largest cluster $n$. Trajectories that pass through configurations with a large value of $\nfree=\nads-n$ are disordered, with many adsorbed subunits not participating in ordered assemblies. In Fig.~\ref{snapsGrowth}, these quantities are shown as a function of time averaged over all simulation trajectories (leading to successful assembly or not), for parameter sets that respectively lead to the en masse mechanism (Fig.~\ref{snapsGrowth}A) and nucleation-and-growth mechanism (Fig.~\ref{snapsGrowth}B). In the first case, there are strong subunit-polyelectrolyte interactions (low ionic strength, $\Csalt{=}100$ mM) and weak subunit-subunit interactions ($\ess{=}3\kt$). Subunits therefore initially adsorb nonspecifically and form only transient subunit-subunit interactions, leading to a rapid rise in $\nads$ with $n\approx 0$. Once enough subunits are adsorbed ($\sim12$ around an optimal-length polyelectrolyte for this model with a 12-subunit capsid), a cooperative fluctuation in subunit configurations eventually leads to a stable nucleus and then rapid completion of the ordered capsid geometry. Since this nucleation process is stochastic, there is a distribution of waiting times and thus a more gradual increase in the average cluster size $n$ (see Fig.~\ref{examples}). In the nucleation-and-growth case, on the other hand, the subunit-polyelectrolyte interactions are weak ($\Csalt{=}300$ mM) and the subunit-subunit interactions are strong ($\ess{=}6\kt$). There is limited nonspecific subunit adsorption onto the polyelectrolyte, adsorbed subunits form relatively strong associations, and thus $\nads$ and $n$ increase nearly in lockstep. Snapshots from typical trajectories for each of these parameter sets are shown in Fig.~\ref{snapsGrowth}A.

\begin{figure} [bt]
\centering{\includegraphics[width=0.7\columnwidth]{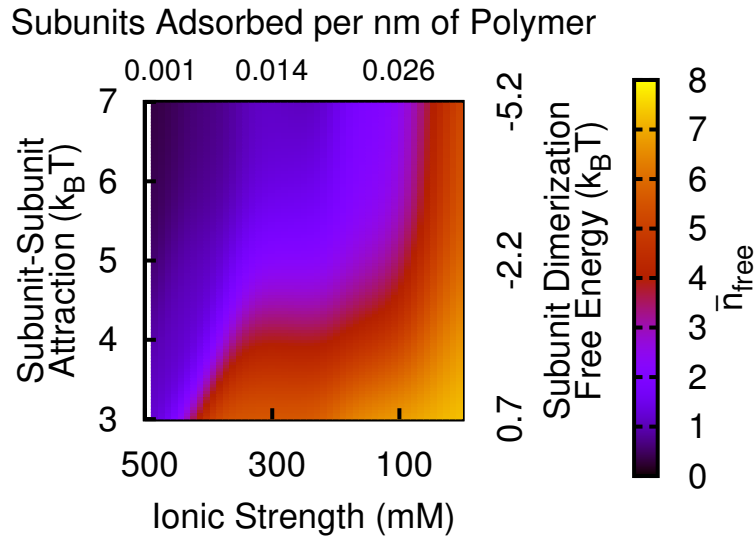}}
\caption{Dependence of the assembly mechanism on the subunit-subunit attraction strength $\ess$ and the ionic strength $\Csalt$. The assembly order parameter $\nfreebar$ (the average number of adsorbed subunits not in the largest partial capsid) is shown as a function of parameter values.  As described in the text, $\nfreebar \lesssim 2$ corresponds to ordered nucleation-and-growth assembly pathways, whereas larger values correspond to disordered pathways.  The alternate y-axis shows the subunit-subunit dimerization free energy $\gss$ corresponding to $\ess$ (see section \ref{sec:freeenergy}) and the alternate x-axis shows the linear adsorption density of subunits in the absence of assembly (see Fig. \ref{affinity} below).
\label{order}
}
\end{figure}

To visualize the degree of order as a function of parameter values, we define a trajectory-averaged order parameter $\nfreebar$, which is $\nfree$ averaged over all configurations with $4\leq n \leq6$ and over all trajectories at a given parameter set\cite{Elrad2008}. Large values of this parameter ($\nfreebar\gtrsim5$) indicate the \textit{en masse} mechanism, while small values ($\nfreebar\lesssim 2$) indicate the nucleation-and-growth mechanism. As shown in Fig.~\ref{order}, the degree of order generally increases with ionic strength and subunit-subunit interaction strength, with the most ordered assembly occurring at $\Csalt=500$ mM (where on average fewer than one subunit is adsorbed nonspecifically) and $\ess\geq6\kt$.  However, notice that for $\ess=3$ assembly is always disordered; for such weak subunit-subunit interactions the critical nucleus size is large and a high density of adsorbed subunits is required to achieve nucleation. On the other hand, for moderate subunit-subunit interactions we do not observe the extreme limit of the en masse mechanism even for low ionic strength. Though a low ionic strength drives strong nonspecific subunit adsorption, absorbed subunits collide frequently due to cooperative polymer motions and subunit sliding along the polymer \cite{Hu2007, Elrad2010}. For $\ess>3$ absorbed subunits therefore achieve nucleation before nonspecific absorption has time to saturate.

The nucleation-and-growth trajectories can be further classified based on the relative timescales of nucleation and growth. When nucleation is slow compared to growth (Figs.~\ref{snapsGrowth}C and \ref{twoState}), the reaction is effectively two-state --- each critical nucleus rapidly proceeds to completion, leading to low concentrations of intermediates which are essentially undetectable in a bulk experiment. When nucleation and growth timescales are comparable, multiple capsids within a spatial region can be assembling simultaneously, and thus potentially could be detected in bulk experiments.  Below, we consider whether it is possible to experimentally distinguish between the latter case, ordered assembly with rapid nucleation, and \textit{en masse} assembly pathways characterized by disordered intermediates.

\begin{figure}
\centering{\includegraphics[width=0.75\columnwidth]{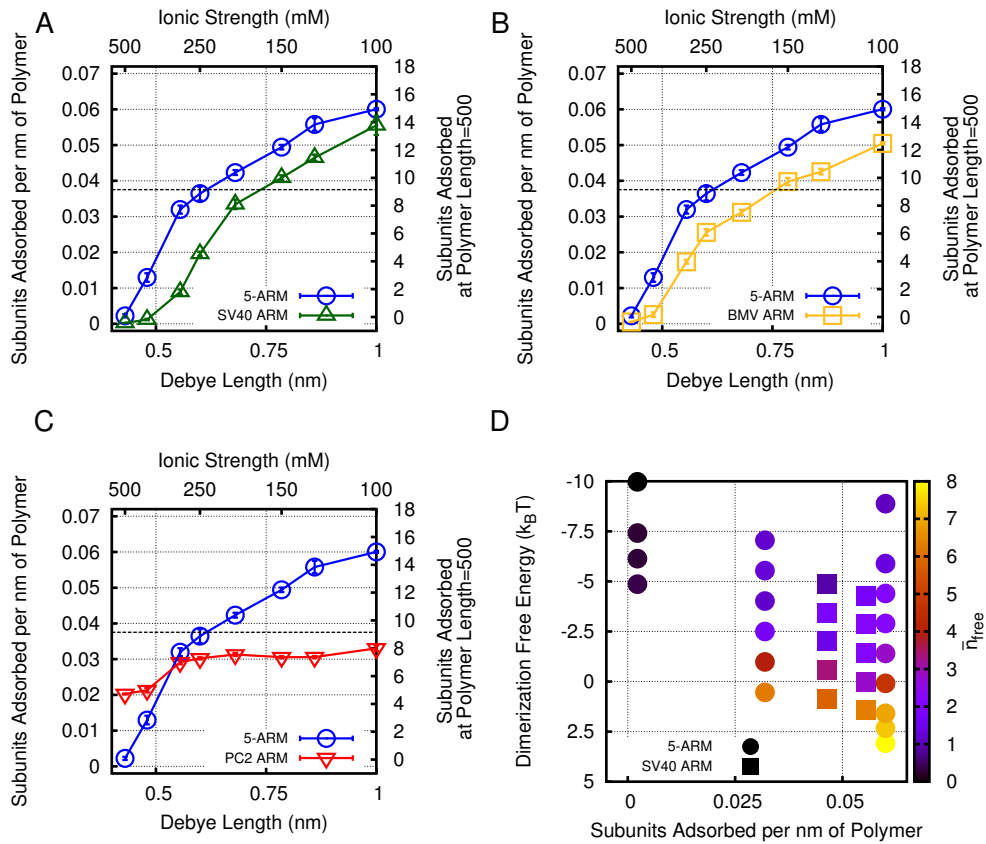}}
\caption{Average number of subunits adsorbed to polymer (in absence of assembly), depending on Debye Length and ARM sequence. Comparison between simple, +5-ARM and SV40 {\bf (A)}, BMV {\bf (B)}, and PC2 {\bf (C)}. {\bf (D)} $\nfreebar$ values obtained during assembly simulations are plotted for varying values of $\ceq$ and subunit-subunit dimerization free energy for our simple 5-ARM assembly model (circles) and the SV40 model (squares). ARM net charge and total length: SV40 $+6$/22, BMV $+9$/44, PC2 $+22$/43.
\label{affinity}
}
\end{figure}

{\bf Biological ARM sequences.} The assembly simulations described above consider a simplified peptide ARM, which contains only 5 positive ARM charges (Fig.~\ref{schematic}). However, our previous work demonstrated that the ARM amino acid sequence (specifically the charges) can significantly affect binding to a polyelectrolyte \cite{Perlmutter2013}. Furthermore, results from a more simplified model suggested that the polyelectrolyte binding affinity is a determining factor for the assembly mechanism \cite{Elrad2010}. To test this hypothesis and to place our results in the context of specific viruses, we calculated the polyelectrolyte binding affinity for the simplified ARM and three ARM sequences from biological viruses (SV40, BMV, PC2), with each amino acid classified as neutral, cationic, or anionic.  These calculations were performed by setting the subunit-subunit interaction strength to zero and recording the average number of adsorbed subunits at varying $\Csalt$ (or Debye length $\ld$).
The measured equilibrium linear densities of adsorbed subunits, $\ceq$, are shown as a function of $\Csalt$ in Fig.~\ref{affinity}A-C.  In all cases the linear density increases monotonically with decreasing $\Csalt$; saturating at a maximum density.
The simplified ARM has the largest binding affinity, despite having the lowest net positive charge ($+5$, 0 neutral segments) of the four species considered. Comparison with SV40 ($+6$, 14 neutral segments) and BMV ($+9$, 33 neutral segments) illustrates the expected result that that neutral segments decrease the binding affinity, particularly at high salt. The PC2 subunits, with net charge $+22$ demonstrate markedly different behavior, with significant subunit absorption at the highest salt concentration simulated (500 mM), but saturating at about 300 mM ionic strength due to subunit-subunit charge repulsions. Variations in adsorption density with subunit concentration are shown in Fig.~\ref{affinityDensity}.

In Fig.~\ref{affinity}D we connect these measures of adsorption with assembly mechanism by plotting the assembly order parameter $\nfreebar$ as a function of $\ceq$ (controlled by varying $\Csalt$), for several values of the subunit dimerization free energy (determined by $\ess$ and $\Csalt$) for the 5-ARM and SV40 models. Plotting against $\ceq$ rather than $\Csalt$ ($\ld$) allows us to overlay data from these two models while accounting for the differences in affinity due to ARM sequence described above. In support of the proposed link between binding affinity and assembly mechanism, we find rough agreement in the measured assembly order parameters between the two models.
The results indicate that ARM sequence can significantly influence the assembly mechanism. For example, if we define $\nfreebar\leq2$ as the nucleation-growth mechanism, Fig.~\ref{affinity}D indicates that nucleation-growth occurs for $\ceq\leq\cstar$ with the threshold value $\cstar=0.0375$ for $\ess\geq2.5\kt$. From Fig.~\ref{affinity}A-C, we can then identify the threshold values of ionic strength $\Istar$, above which the nucleation-growth mechanism will occur: $\Istar \approx 300$ mM for the 5-ARM and $\Istar \approx 175$ mM for the BMV and SV40 models; while PC2 is below the threshold value for all salt concentrations considered. This allows us to predict, for example, that recent experiments on SV40 assembly (at $\Csalt{=}250$mM and observed strong subunit-subunit attraction) would have a very low $\nfreebar$ ($\sim1$), which is consistent with SAXS observations \cite{Kler2012}.


\subsection{Experimental Observables}
 We now seek to provide experimental signatures through which the capsid assembly pathways discussed above can be distinguished. We focus on two experimental techniques which have recently been applied to respectively probe the formation of individual capsids and bulk assembly kinetics. 
 
 {\bf smFCS measurements on individual capsids can distinguish assembly mechanisms.} Borodavka et al. \cite{Borodavka2012} used single molecule fluorescence correlation spectroscopy (smFCS) to monitor the hydrodynamic radii $\Rh$ of nucleocapsid complexes during assembly of MS2 and STNV capsid proteins around cognate RNA or non-cognate RNA. Assembly around cognate RNA was characterized by either constant $\Rh$ or, in some trajectories, a collapsed complex followed by gradual increase until reaching the size of an assembled capsid. In contrast, assembly around non-cognate RNA led to an increase in $\Rh$ before eventually decreasing to the size of the capsid. The difference between these two behaviors can be attributed to sequence-specific `packaging signals' on cognate RNA that interact with the capsid proteins. In this article we do not consider the effect of packaging signals (these will be considered in a subsequent article); instead, we consider whether the pathways described in the previous section can be distinguished by this experimental technique.

\begin{figure}
\centering{\includegraphics[width=0.9\columnwidth]{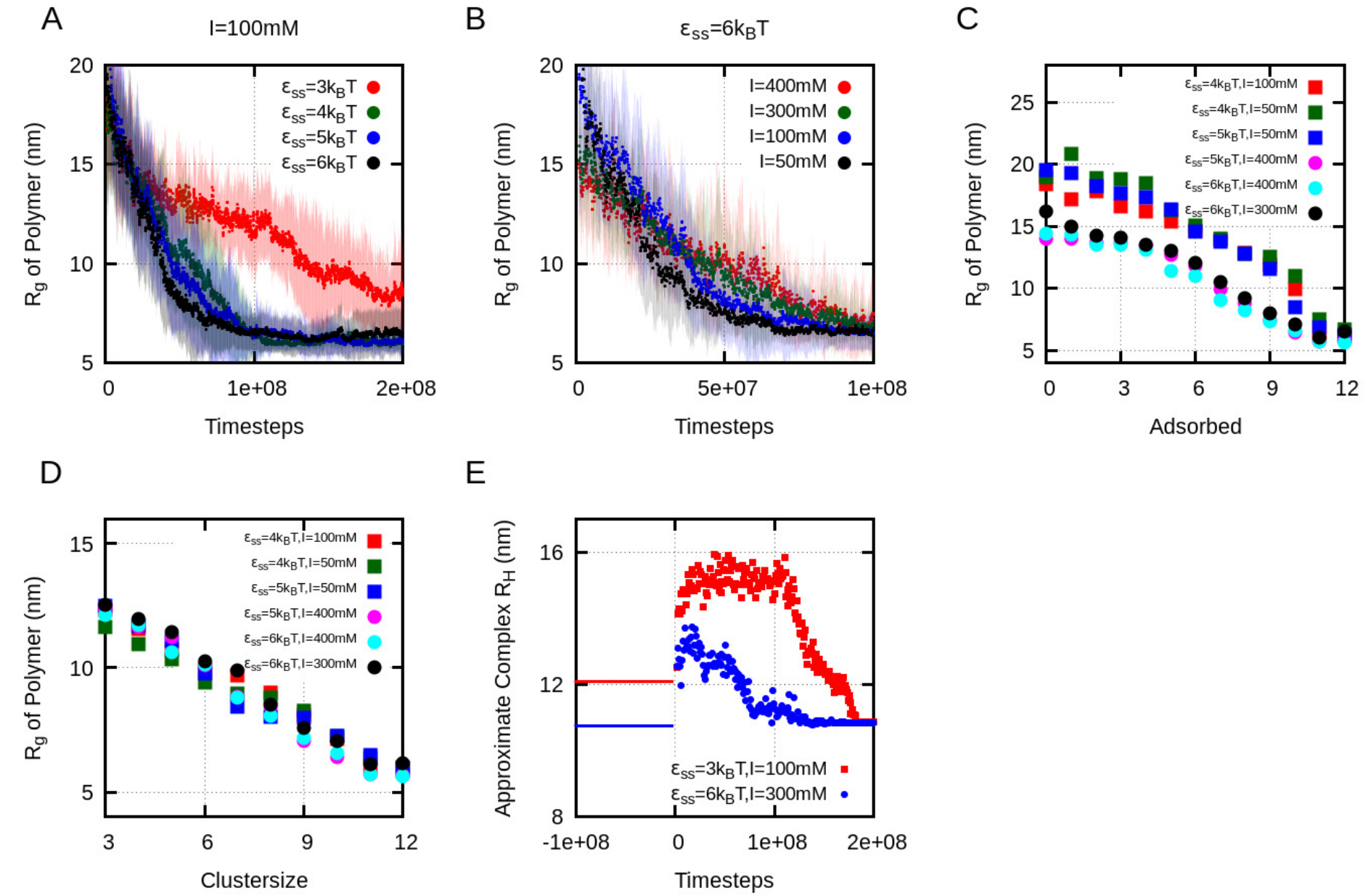}}
\caption{{\bf(A),(B)} Average polymer radius of gyration during assembly as a function of varying $\ess$ with constant $\Csalt$ {\bf (A)} and varying $\Csalt$ with constant $\ess$ {\bf (B)}. {\bf (C),(D)} Average polymer radius of gyration during assembly as a function of number of adsorbed subunits $\nads$ {\bf (C)} and size of the largest cluster $n$ {\bf (D)} for parameters that lead to relatively disordered (squares) and ordered (circles) pathways. {\bf (E)} Approximate average radius of hydration for representative ordered and disordered trajectories.
\label{radius}
}
\end{figure}

We estimated the hydrodynamic radii $\Rh$ for polymer-subunit intermediates using the program HYDROPRO, which has been shown to accurately predict $\Rh$ for large protein complexes \cite{Ortega2011}. The resulting $\Rh$ are shown during representative en masse and ordered assembly trajectories in Fig.~\ref{radius}E.  We see that the complex $\Rh$ first increases as subunits adsorb onto the complex and then decreases as subunits assemble (Fig.~\ref{radius}E). However, the en masse mechanism leads to a much larger and longer duration increase in $\Rh$, due to the extensive and long-lived disordered adsorption of unassembled subunits. The difference in $\Rh$ between \textit{en masse} and ordered trajectories is conserved across their respective parameter ranges (see SI Fig.~\ref{radiusSupplement} for other trajectories), and also occurs for assembly trajectories around the model nucleic acid with intramolecular base- pairing developed in Ref. \cite{Perlmutter2013} (see SI Fig.~\ref{radiusSupplement}C,F). These results suggest that smFCS can distinguish among the classes of assembly pathways observed in our simulations. They are consistent with an interpretation of the Borodavka et al. \cite{ Borodavka2012} results in which assembly around the non-cognate RNA proceeds via the disordered mechanism while packaging signals lead to an ordered mechanism.

To further characterize the differences in polymer conformations between disordered and ordered assembly pathways, we show the polymer radius of gyration, $\Rg$, during assembly trajectories in Fig.~\ref{radius}. In contrast to $\Rh$, contributions of the capsid proteins are not included in $\Rg$.  While the results in Fig.~\ref{radius} are averaged over multiple trajectories, example individual trajectories are shown in SI Fig.~\ref{examples}.  In all cases of successful assembly, the polymer is gradually compacted from its free size into its encapsidated size. However, at fixed $\Csalt{=}100$mM, the average rate of compaction increases with $\ess$, with a dramatic increase in rate for $\ess>3 \kt$ (Fig.~\ref{radius}A).  Similarly, decreasing $\Csalt$ increases the rate of compaction (Fig.~\ref{radius}B).  Notice that the rate of polymer compaction is not determined by the assembly mechanism --- increased order correlates with faster compaction in Fig.~\ref{radius}A but with slower compaction in Figure~\ref{radius}B.  When the $\Rg$ is plotted as a function of number of adsorbed subunits ($\nads$), the en masse and ordered pathways clearly split into two groups (Fig.~\ref{radius}C). However, this distinction disappears when $\Rg$ is plotted as a function of the number of subunits in the largest cluster ($n$, Fig.~\ref{radius}D). Taken together, these data demonstrate that polymer compaction is driven by adsorbed subunits forming ordered intermediates, with the rate of compaction consequently mirroring the rate of capsid assembly.

{\bf trSAXS measurments of bulk assembly kinetics can distinguish assembly mechanisms.} While smFCS can detect individual assembly intermediates, Kler et al. \cite{Kler2012,Kler2013} recently used time resolved small angle x-ray scattering (trSAXS) to elucidate structures of assembling capsids in bulk.  They found that the SAXS profiles at all time points could be decomposed into scattering contributions from the unassembled and complete components, suggesting that assembly proceeded by an effectively two-state reaction with undetectable levels of intermediates (i.e. the nucleation-and-growth pathway with relatively slow nucleation).  While it is evident that profiles from profiles from a two-state reaction can be decomposed in this way, we investigated the extent to which SAXS profiles from the other pathway classes (en masse or nucleatation-and-growth with rapid nucleation) can be distinguished from the two-state case.

\begin{figure}
\centering{\includegraphics[width=0.8\columnwidth]{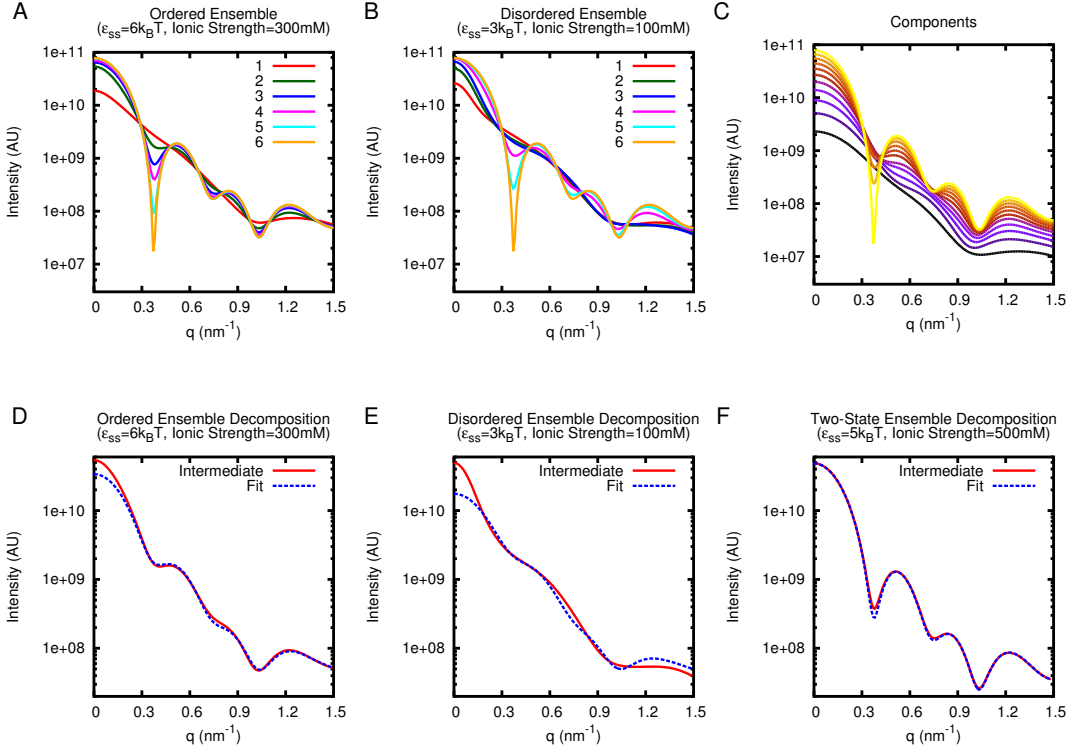}}
\caption{SAXS profiles for {\bf (A)} the nucleation-and-growth (ordered) and {\bf (B)} the \textit{en masse} (disordered) assembly mechanisms, at indicated parameter values. Simulations were divided into six segments of equal length to describe time evolution from beginning (1) to completion (6). {\bf (C)} Scattering of subunit clusters from 1 subunit (black) to a 12 subunit capsid (yellow). {\bf (D,E,F)} Best fit to SAXS profiles at an intermediate time (line 2 in {\bf (A,B)}) assuming two-state kinetics (a linear combination of complete capsid and unassembled subunits) for ordered {\bf (D)}, disordered {\bf (E)}, and ordered with rate-limiting nucleation {\bf (F)}.
\label{saxs}
}
\end{figure}

First, Fig.~\ref{saxs}A,B shows SAXS profiles calculated (using CRYSOL \cite{Svergun1995}) from simulation snapshots along ensembles of \textit{en masse} and ordered assembly trajectories. For each parameter set, SAXS profiles are averaged over 6 time windows. In both cases, the first profile is dominated by scattering from free subunits (Fig.~\ref{saxs}C) and the final time segment shows clear minima and maxima corresponding to the complete capsid (Fig.~\ref{saxs}C). For comparison, Fig.~\ref{saxs}C presents the SAXS profiles for ordered subunit clusters ranging in size from a single subunit (black) to a complete capsid (yellow). As the capsid grows distinct minima and maxima appear and become more pronounced. We note that the positions of the minima and maxima in the complete model capsid are similar to those observed experimentally for SV40 \cite{Kler2012}.

To test the extent to which these trajectories can be distinguished from the two-state case, we attempted to fit SAXS profiles using a linear combination of the profiles for unassembled polymer and subunits and the complete capsid. The resulting fits for the second fifth of the trajectory (where intermediates are most plentiful) are shown in Figs.~\ref{saxs}D--F. At this stage, the ordered systems contain mostly incomplete subunit clusters (6-11 subunits), while the disordered simulations contain mostly adsorbed but unassembled subunits. For the ordered simulations, we find that the fit reproduces all of the features of the intermediates, except at low q. In contrast, the intermediates in the disordered trajectory display a shoulder at $q\sim0.3nm^{-1}$ that is not observed in any of the unassembled or assembled components. This shoulder is a distinct feature of the disordered intermediate, and thus could be used to identify this class of pathways. Finally, as expected, SAXS profiles from trajectories at parameter sets (Fig.~\ref{snapsGrowth}C) which lead to two-state kinetics are very well fit by a linear combination of polymer/subunit and complete capsid (Fig.~\ref{saxs}F).

A significant distinction between the SAXS profiles is that the ordered pathways lead to an  isosbestic point at $q\sim0.3nm^{-1}$ (as seen in SV40 experiments\cite{Kler2012}), whereas the disordered pathways do not. The presence of an isosbestic point is frequently assumed to indicate two-state behaviour; however, it occurs in our data for ordered trajectories even when the reaction is far from two-state due to rapid nucleation. In these cases the isosbestic point appears due to due to the similarity in scattering from the ordered intermediates and the complete capsid. This suggests that an isosbestic point may distinguish ordered from disordered assemblies, but is less sensitive to the extent to which the reaction kinetics can be approximated as two-state (i.e., how undetectable the intermediate concentrations are).


\section{Discussion}

Our simulations demonstrate that capsid assembly around a NA or other cargo can proceed through two mechanisms, and that which mechanism occurs depends on the relative strengths of protein-protein and protein-cargo interactions. The assembly mechanism can be rationally designed by tuning these interactions through solution conditions or mutagenesis of capsid protein-NA binding domains. However, because productive assembly requires weak interactions, the parameters must be tuned within relatively narrow ranges, and alterations which increase (decrease) the strength of one type of interaction must be compensated by a decrease (increase) in the other. Our results suggest that the subunit-cargo dissociation constant is an important parameter, whose value might be used to map specific viruses onto our phase diagrams (Figs. \ref{outcome}, \ref{yield} ,\ref{order}), although experimental tests of this capability are required. 
Finally, we have shown that the nature of assembly pathways can be inferred from the results of recently developed techniques to monitor the assembly of individual capsids or bulk assembly kinetics.

Our simulations predict that a single viral species can be driven to assemble via different mechanisms \textit{in vitro} by changing solution conditions. In particular, under a constant subunit-subunit binding energy ($\ess \sim 5-6\kt$) robust assembly occurs for a range of solvent conditions, with highly ordered or disordered assemblies occurring depending on salt concentration ($\Csalt \in [50,400]$ mM). To our knowledge, this prediction has not yet been realized experimentally, although the signatures of the two classes of assembly pathways have been seen in experiments on different viruses and/or different cargoes \cite{Kler2012, Garmann2013, Cadena-Nava2012, Malyutin2013, Borodavka2012}.

One recent experimental study sought to test the role of of $\Csalt$ and subunit-subunit attractions (controlled by solution $p$H) on \textit{in vitro} assembly of CCMV around RNA \cite{Garmann2013}. In some regards, these experiments mirror our observations, with malformed capsids observed for strong subunit-subunit attractions, disordered products for weak subunit-subunit attractions, and well-formed capsids at intermediate conditions. However, robust assembly was only observed for a two-step process: first $\Csalt$ was reduced (from $1M$) to drive RNA-subunit interactions, and then secondly $p$H was reduced to activate subunit-subunit attractions.  The resulting assembly pathways resemble the \textit{en masse} mechanism described here. On the other hand, one-step assembly led to malformed particles even at moderate $p$H, suggesting an inability to assemble through a nucleation-and-growth mechanism in these experiments. In our simulations, this outcome would only occur at high salt concentrations (e.g. $\Csalt\sim400-500$mM, see Figs.~\ref{outcome}, \ref{yield}), where the narrow range of $\ess$ leading to successful assembly indicates that parameters must be finely tuned. Reproducing such a lack of successful assembly at moderate salt concentrations would require a reduction in the orientation-dependence of the subunit-subunit interaction potential (see Methods), or introduction of additional factors as discussed below. Experiments in which solution conditions are changed for other viruses which do undergo one step assembly (e.g. SV40 \cite{Kler2012,Kler2013}), may elucidate which of these possibilities to pursue and would directly test our prediction that the assembly mechanism can be controlled by solution conditions.

Understanding capsid assembly mechanisms and their location within the assembly phase diagram has important implications for the design of antiviral agents. 
 As one example, we consider the recently developed class of HBV inhibitors based on phenylpropenamides \cite{Katen2010, Katen2013}, which act by {\bf increasing} the strength of subunit-subunit interactions,  driving subunits to assemble in the absence of their genome and thus increasing the generation of empty, non-infective capsids \cite{Katen2010, Katen2013}. Comparing Figs.~\ref{yield} and \ref{order} shows that a virus that undergoes ordered assembly (e.g. $\ess \sim 6-7$, $\Csalt \sim 300$ mM), sits close to parameters that support empty capsid assembly, which are demarcated by dashed lines in Fig.~\ref{yield}. Thus only a small increase in subunit-subunit interactions is required to trigger unproductive RNA-free assembly. In contrast, a much larger perturbation would be required to achieve empty capsid assembly for a virus that assembles via the en masse mechanism.

{\bf Outlook.}
We have described two classes of assembly pathways and several distinct failure modes (Fig.~\ref{outcome}) that arise when assembly is driven by nonspecific electrostatic subunit-cargo interactions. Our phase diagrams can serve as a starting point to understand how virus-specific features, such as packaging signals \cite{Stockley2013}, allosteric NA-induced \cite{Kler2013,Stockley2007} or `autosteric' protein-induced conformational changes \cite{Caspar1980}, base pairing-induced NA-structure \cite{Perlmutter2013,Yoffe2008,Borodavka2012}, or subcellular localization \cite{Bamunusinghe2011} can robustly avoid failure modes amidst the crowded intracellular milieu while enabling selective assembly around the viral genome \cite{Routh2012,Ford2013,Rao2006}. For example, allosteric or autosteric conformational changes may allow for strong subunit-subunit interactions on the NA while avoiding the off-cargo assembly we observe at large $\ess$. Systematically studying how these additional factors expand (or contract) regions of parameter space that lead to successful assembly will ultimately reveal how viruses have optimized their structures and interactions for robust assembly \textit{in vivo} and how their assembly \textit{in vivo} or \text{in vitro} can be manipulated for biomedical or materials science applications.

\section{Methods}
{\bf Model.} We have recently a presented a complete description of our model system, which we summarize here briefly \cite{Perlmutter2013} and in the SI. Our model subunits are based upon that previous used to simulate assembly of empty capsids \cite{Wales2005, Fejer2009, Johnston2010} which we extended previously to model assembly around cargo \cite{Perlmutter2013}. The pseudoatoms in the capsid subunit model are illustrated in Fig.~\ref{schematic}. Subunit assembly is mediated through an attractive Morse potential between Attractor (`A') pseudoatoms located at each subunit vertex. The Top (`T') pseudoatoms interact with other `T' psuedoatoms through a potential consisting of the repulsive term of the Lennard-Jones (LJ) potential, the radius of which is chosen to favor a subunit-subunit angle consistent with a dodecahedron (116 degrees). The Bottom (`B') pseudoatom has a repulsive LJ interaction with `T' pseudoatoms, intended to prevent `upside-down' assembly. The `T', `B', and `A' pseudoatoms form a rigid body \cite{Wales2005, Fejer2009, Johnston2010}. See Refs. \cite{Schwartz1998, Hagan2006, Hicks2006, Nguyen2007, Wilber2007, Nguyen2008, Nguyen2009, Johnston2010, Wilber2009a, Wilber2009, Rapaport1999, Rapaport2004, Rapaport2008, Hagan2008, Elrad2010, Hagan2011, Mahalik2012, Levandovsky2009} for related models. 

To model electrostatic interaction with a negatively charged NA or polyelectrolyte we extend the model as follows. Firstly, to better represent the capsid shell we add a layer of `Excluder' pseudoatoms which have a repulsive LJ interaction with the polyelectrolyte and the ARMs. Each ARM is modeled as a bead-spring polymer, with one bead per amino acid. The `Excluders' and first ARM segment are part of the subunit rigid body. ARM beads interact through repulsive LJ interactions and, if charged, electrostatic interactions modelled by a Debye-Huckel potential. We note that repulsion between subunits due to the positive charges does affect the magnitude of the subunit-subunit interaction. Previously, we estimated this repulsion contributes $1\kt$ to the dimerization free energy \cite{Perlmutter2013}. This contribution (and entropic terms) are not included in $\ess$, which is the magnitude of the Morse potential depth. See SI section \ref{sec:freeenergy}) for a discussion of binding free energies.

{\bf Simulations and units.} Simulations were performed with the Brownian Dynamics algorithm of HOOMD, which uses the Langevin equation to evolve positions and rigid body orientations in time \cite{Anderson2008, nguyen2011, LeBard2012}. Simulations were run using a set of fundamental units. The fundamental energy unit is selected to be $\eunit\equiv1 \kt$. The unit of length $\dunit$ is set to the circumradius of a pentagonal subunit, which is taken to be $1\dunit\equiv5$ nm so that the dodecahedron inradius of $1.46\dunit=7.3$ nm gives an interior volume consistent with that of the smallest $T{=}1$ capsids. Assembly simulations were run at least 10 times for each set of parameters, each of which were concluded at either completion, persistent malformation, or $2\times10^8$ time steps. For all dynamics simulations there were 60 subunits with $\mbox{box size} {=} 200 \times 200 \times 200$ nm, resulting in a concentration of  $12 \mu$M. 

{\bf SAXS profile and hydrodynamic radius estimations.} Small Angle X-ray Scattering (SAXS) analysis was performed using CRYSOL \cite{Svergun1995}. For this analysis the all-atom structure of an SV40 pentameric subunit \cite{Stehle1996} was aligned into the position of each coarse-grained subunit and the polymer was replaced with a carbon chain. We note that this entails significant simplification: segments of the protein which were not resolved in the crystal structure were not reconstructed and there is no optimization of structure at the all-atom resolution. We believe this approximation is suitable, given that our analysis is limited to the X-ray scattering profile in the small angle regime, which reflects $\sim$nm scale structural features. Fitting of the scattering profiles was performed using least squares fitting. Hydrodynamic radius analysis was performed using HYDROPRO \cite{Ortega2011}. This program is capable of calculating the hydrodynamic radius of large protein complexes. In order to perform this analysis, we treated the synthetic polymer as an amino acid chain. Though this is a gross approximation, it has a negligible effect: removing the polymer entirely does not change the $\Rh$ trend and only alters the magnitude by $\sim1\%$. Simulations were visualized using VMD \cite{Humphrey1996}.

\section{Acknowledgments}
We gratefully acknowledge Adam Zlotnick for insightful discussion and critical reading of the manuscript. This work was supported by Award Number R01GM108021 from the National Institute Of General Medical Sciences. Computational resources were provided by the NSF through XSEDE computing resources (Longhorn, Maverick, Condor, and Keeneland) and the Brandeis HPCC which is partially supported by NSF-MRSEC-0820492.

\bibliographystyle{plain}
\bibliography{all-references}

\pagebreak
\setcounter{figure}{0}
\renewcommand{\thefigure}{S\arabic{figure}}
\section{Supporting Information}

\subsection{Model potentials and parameters}
\label{sec:potentials}
The model details are described in Ref. \cite{Perlmutter2013}; we summarize them here for the convenience of the reader. In our model, all potentials can be decomposed into pairwise interactions. Potentials involving capsomer subunits further decompose into pairwise interactions between their constituent building blocks -- the excluders, attractors, `Top' and `Bottom', and ARM pseudoatoms. It is convenient to write the total energy of the system as the sum of 6 terms: a capsomer-capsomer $U\sub{cc}{}$ part (which does not include interactions between ARM pseudoatoms), capsomer-polymer $U\sub{cp}{}$, capsomer-ARM $U\sub{ca}{}$, polymer-polymer $U\sub{pp}{}$, polymer-ARM $U\sub{pa}{}$, and ARM-ARM $U\sub{aa}{}$ parts. Each is summed over all pairs of the appropriate type:
\begin{align}
U = & \sum_{\mathrm{cap\ }{i}} \sum_{\mathrm{cap\ }{j < i}} U\sub{cc}{}
  + \sum_{\mathrm{cap\ }{i}} \sum_{\mathrm{poly\ }{j}} U\sub{cp}{}
  + \sum_{\mathrm{cap\ }{i}} \sum_{\mathrm{ARM\ }{j}} U\sub{ca}{} + \nonumber \\
  & \sum_{\mathrm{poly\ }{i}} \sum_{\mathrm{poly\ }{j < i}} U\sub{pp}{}
  + \sum_{\mathrm{poly\ }{i}} \sum_{\mathrm{ARM\ }{j}} U\sub{pa}{}
  + \sum_{\mathrm{tail\ }{i}} \sum_{\mathrm{ARM\ }{j < i}} U\sub{aa}{}
\end{align}
where $\sum_{\mathrm{cap\ }{i}} \sum_{\mathrm{cap\ }{j < i}}$ is the sum over all distinct pairs of capsomers in the system, $\sum_{\mathrm{cap\ }{i}} \sum_{\mathrm{poly\ }{j}}$ is the sum over all capsomer-polymer pairs, etc.

The capsomer-capsomer potential $U\sub{cc}{}$ is the sum of the attractive interactions between complementary attractors, and geometry guiding repulsive interactions between `Top' - `Top' pairs and `Top' - `Bottom' pairs. There are no interactions between members of the same rigid body, but ARMs are not rigid and thus there are intra-subunit ARM-ARM interactions. Thus, for notational clarity, we index rigid bodies and non-rigid pseudoatoms in Roman, while the pseudoatoms comprising a particular rigid body are indexed in Greek. E.g., for capsomer $i$ we denote its attractor positions as $\{\bm{a}_{i\alpha}\}$ with the set comprising all attractors $\alpha$, its `Top' positions $\{\bm t_{i\alpha}\}$, and its `Bottom' positions $\{\bm b_{i\alpha}\}$. The capsomer-capsomer interaction potential between two capsomers $i$ and $j$ is then defined as:

\begin{align}
\label{Ucc}
U\sub{cc}{}&(\{\bm a_{i\alpha}\}, \{\bm t_{i\alpha}\}, \{\bm b_{i\alpha}\} , \{\bm a_{j\beta}, \{\bm t_{j\beta}\}, \{\bm b_{j\beta}\})  = \nonumber \\
&\sum_{\alpha,\beta}^{N\sub{t}{}} \varepsilon \LJ{} \left(
\left|\bm{t}_{i\alpha} - \bm{t}_{j\beta} \right|,
\ \sigma\sub{t}{} \right) + \nonumber \\
&\sum_{\alpha,\beta}^{N\sub{b}{},N\sub{t}{}} \varepsilon \LJ{} \left(
\left| \bm{b}_{i\alpha} - \bm{t}_{j\beta} \right|,
\ \sigma\sub{b}{} \right) + \nonumber \\
&\sum_{\alpha,\beta}^{N\sub{a}{}} \varepsilon \Morse{} \left(
\left|\bm{a}_{i\alpha} - \bm{a}_{j\beta} \right|,
\ r\sub{0}{}, \varrho, \rcut \right)
\end{align}
where $\varepsilon$ is an adjustable parameter which both sets the strength of the capsomer-capsomer attraction at each attractor site and scales the repulsive interactions which enforce the dodecahedral geometry. $N\sub{t}{}$, $N\sub{b}{}$, and $N\sub{a}{}$ are the number of `Top', `Bottom', and attractor pseudoatoms respectively in one capsomer, $\sigma\sub{t}{}$ and $\sigma\sub{b}{}$ are the effective diameters of the `Top' -- `Top' interaction and `Bottom' -- `Top' interactions, which are set to 10.5 nm and 9.0 nm respectively throughout this work, $r\sub{0}{}$ is the minimum energy attractor distance, set to 1 nm, $\varrho$ is a parameter determining the width of the attractive interaction, set to 2.5, and $\rcut$ is the cutoff distance for the attractor potential, set to 10.0 nm.

The function $\LJ{}$ is defined as the repulsive component of the Lennard-Jones potential shifted to zero at the interaction diameter:
\begin{equation}
\LJ{}(x,\sigma) \equiv
\left\{  \begin{array}{ll}
	\left(\frac{\sigma}{x}\right)^{12} -1 & : x < \sigma \\
	0 & : \mathrm{otherwise}
\end{array} \right.
\label{eq:LJ}
\end{equation}
The function $\Morse{}$ is a Morse potential:
\begin{equation}
\Morse{}(x,r\sub{0}{},\varrho) \equiv
\left\{  \begin{array}{ll}
\left(e^{\varrho\left(1-\frac{x}{r\sub{0}{}}\right)} - 2 \right)e^{\varrho\left(1-\frac{x}{r\sub{0}{}}\right)} & : x < \rcut \\
 0 & : \mathrm{otherwise}
\end{array} \right.
\label{eq:Morse}
\end{equation}

The capsomer-polymer interaction is a short-range repulsion that accounts for excluded-volume. For capsomer $i$ with excluder positions $\{\bm x_{i\alpha}\}$ and polymer subunit $j$ with position $\bm R_j$, the potential is:
\begin{eqnarray}
\label{Ucp}
U\sub{cp}{}(\{\bm x_{i\alpha}\}, \bm R_j) &=&
    \sum_{\alpha}^{N\sub{x}{}} \LJ{} \left(
    | \bm{x}_{i\alpha} - \bm R_j |,
    \sigma\sub{xp}{}\right)
\end{eqnarray}
where $N_\text{x}$ is the number of excluders on a capsomer and $\sigma\sub{xp}{} = 0.5(\sigma_\text{x}+\sigma_\text{p})$ is the effective diameter of the excluder -- polymer repulsion. The diameter of the polymer bead is $\sigma_\text{p}=0.5$ nm and the diameter for the excluder beads is $\sigma_\text{x}=3.0$ nm for the $T{=}1$ model and $\sigma_\text{x}=5.25$ nm for the $T{=}3$ model.

The capsomer-ARM interaction is a short-range repulsion that accounts for excluded-volume. For capsomer $i$ with excluder  positions $\{\bm x_{i\alpha}\}$ and ARM subunit $j$ with position $\bm R_j$, the potential is:
\begin{eqnarray}
\label{UcA}
U\sub{cA}{}(\{\bm x_{i\alpha}\}, \bm R_j) &=&
    \sum_{\alpha}^{N\sub{x}{}} \LJ{} \left(
    | \bm{x}_{i\alpha} - \bm R_j |,
    \sigma\sub{xA}{}\right)
\end{eqnarray}
with $\sigma\sub{xA}{} = 0.5(\sigma_\text{x}+\sigma_\text{A})$ as the effective diameter of the excluder - ARM repulsion with $\sigma_\text{A}=0.5$ nm the diameter of an ARM bead.

The polymer-polymer non-bonded interaction is composed of electrostatic repulsions and short-ranged excluded-volume interactions. These polymers also contain bonded interactions which are only evaluated for segments occupying adjacent positions along the polymer chain and angular interactions which are only evaluated for three sequential polymer segments. As noted in the main text, electrostatics are represented by Debye Huckel interactions.

\begin{align}
\label{Upp}
U\sub{pp}{}(\bm R_i, \bm R_j, \bm R_k)
    &=&
    \left\{
            \begin{array}{l}
                \mathcal{K}_\mathrm{bond}(R_{ij}, \sigma\sub{p}{},k_\mathrm{bond})\\
                \quad:\{i,j\}\ \mathrm{bonded} \\
                \mathcal{K}_\mathrm{angle}(R_{ijk}, k_\mathrm{angle})\\
                \quad: \{i,j,k\}\ \mathrm{angle} \\
                \LJ{}(R_{ij}, \sigma\sub{p}{}) + \mathcal{U}\rsub{DH}(R_{ij},\qp,\qp, \sigma\sub{p}{})\\
                \quad: \{i,j\}\ \mathrm{nonbonded}\\
            \end{array}
        \right.
\end{align}
where $R_{ij} \equiv | \bm R_i - \bm R_j |$ is the center-to-center distance between the polymer subunits, $\qp=-1$ is the valence of charge on each polymer segment, and $\mathcal{U\rsub{DH}}$ is a Debye-Huckel potential smoothly shifted to zero at the cutoff:

\begin{align}
\mathcal{U}\rsub{DH}&(r,q_1,q_2, \sigma{}) \equiv \\
&\left\{  \begin{array}{l}
	\frac{q_1 q_2 l\rsub{b}\ e^{\sigma{}/\lambda\rsub{D}}}{\lambda\rsub{D}+\sigma{}} \left(\frac{e^{-r/\lambda\rsub{D}}}{r} \right)\\
	\quad: x < 2\lambda\rsub{D} \\
	 \frac{(r_{cut}^{2}-r^{2})^{2}(r_{cut}^{2}+2r^{2}-3r_{on}^{2}))}{(r_{cut}^{2}-2r_{on}^{2})^{3}} \frac{q_1 q_2 l\rsub{b}\ e^{\sigma{}/\lambda\rsub{D}}}{\lambda\rsub{D}+\sigma{}}
	\left(\frac{e^{-r/\lambda\rsub{D}}}{r} \right)\\
	\quad: 2\lambda\rsub{D} < x < 3\lambda\rsub{D} \\
	0 \\
	\quad: \mathrm{otherwise}
\end{array} \right.
\end{align}

$\lambda\rsub{D}$ is the Debye length, $l\rsub{b}$ is the Bjerrum length, and $q_1$ and $q_2$ are the valences of the interacting charges.

Bonds are represented by a harmonic potential:
\begin{equation}
    \mathcal{K}_\mathrm{bond}(R_{ij}, \sigma, k_\mathrm{bond})
    \equiv
    	 \frac{k_\mathrm{bond}}{2}(R_{ij}-\sigma)^2
    \label{eq:bonds}.
\end{equation}
Angles are also represented by a harmonic potential:
\begin{equation}
    \mathcal{K}_\mathrm{angle}(R_{ijk}, k_\mathrm{angle})
    \equiv
    	 \frac{k_\mathrm{angle}}{2}(\vartheta_{ijk})^2
\end{equation}
where $\vartheta_{ijk}$ is the angle formed by the sequential polymer units $i,j,k$.

The ARM-ARM interaction is similar to the polymer-polymer interaction, consisting of non-bonded interactions composed of electrostatic repulsions and short-ranged excluded-volume interactions. These ARMs also contain bonded interactions which are only evaluated for segments occupying adjacent positions along the polymer chain:

\begin{eqnarray}
\label{Uaa}
U\sub{aa}{}(\bm R_i, \bm R_j)
    &=&
    \left\{
            \begin{array}{l}
                \mathcal{K}_\mathrm{bond}(R_{ij}, \sigma\sub{a}{},k_\mathrm{bond})\\
                \quad: \{i,j\}\ \mathrm{bonded} \\
                \LJ{}(R_{ij}, \sigma\sub{a}{}) + \mathcal{U}\rsub{DH}(R_{ij},q_i,q_j, \sigma\sub{a}{})\\
                \quad: \{i,j\}\ \mathrm{nonbonded}\\
            \end{array}
        \right.
\end{eqnarray}
where $R_{ij} \equiv | \bm R_i - \bm R_j |$ is the center-to-center distance between the ARM subunits and $q_i$ is the valence of charge on ARM segment $i$.

Finally, the ARM-Polymer interaction is the sum of short-ranged excluded-volume interactions and electrostatic interactions:

\begin{equation}
\label{Upa}
U\sub{pa}{}(\bm R_i, \bm R_j)
    =
	\LJ{}(R_{ij}, \sigma\sub{ap}{}) + \mathcal{U}\rsub{DH}(R_{ij},q_i,q_j, \sigma\sub{ap}{})
\end{equation}

\subsection{Calculation of binding free energy estimates}
\label{sec:freeenergy}
\textit{Subunit-Subunit binding free energy.} Our method of calculating the subunit-subunit binding free energy was described previously \cite{Perlmutter2013} and is similar to that presented in our previous work \cite{Elrad2010, Hagan2011}. Briefly, subunits were modified such that only one edge formed attractive bonds, limiting complex formation to dimers. We measured the relative concentration of dimers and monomers for a range of attraction strengths ($\ess$). The free energy of binding along that interface is then $\gss/\kt =- \ln(c_\text{ss}/\kd)$ with standard state concentration $c_\text{ss}=1$ M and $\kd$ in molar units. We can then correct for the multiplicity of dimer conformations, by adding in the additional term $-T\Delta\scc = \ln(25/2)\kt$, where the five pentagonal edges are assumed to be distinguishable, but complex orientations which differ only through global rotation are not. For subunits which do not contain ARMs, the free energy is well fit by the linear expression $\gss/\kt = -1.5\ess - T\sbb$ where $T\sbb=-5.0\kt$, and these values are used in Figures~\ref{yield} and ~\ref{order}. Calculations were also performed with ARMs at varying $\Csalt$. For $\Csalt{=}100$ mM, $\gss$ increases by $\sim0.5\kt$ due to ARM-ARM repulsion for the simple 5-ARM model. For the SV40 ARM, $\gss$ increases by $\sim2\kt$ at $\Csalt{=}100$ mM.

\begin{figure}[p]
\centering{\includegraphics[width=0.35\columnwidth]{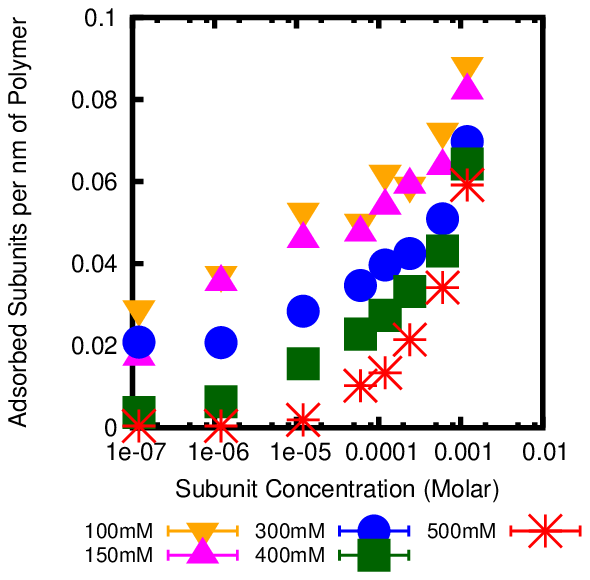}}
\caption{Linear adsorption density of subunits on the polyelectrolyte (per nm), as a function of subunit concentration for varying salt concentrations, for simulations with no subunit-subunit attraction ($\ess{=}0\kt$).}
\label{affinityDensity}
\end{figure}

\begin{figure}[p]
\centering{\includegraphics[width=0.95\columnwidth]{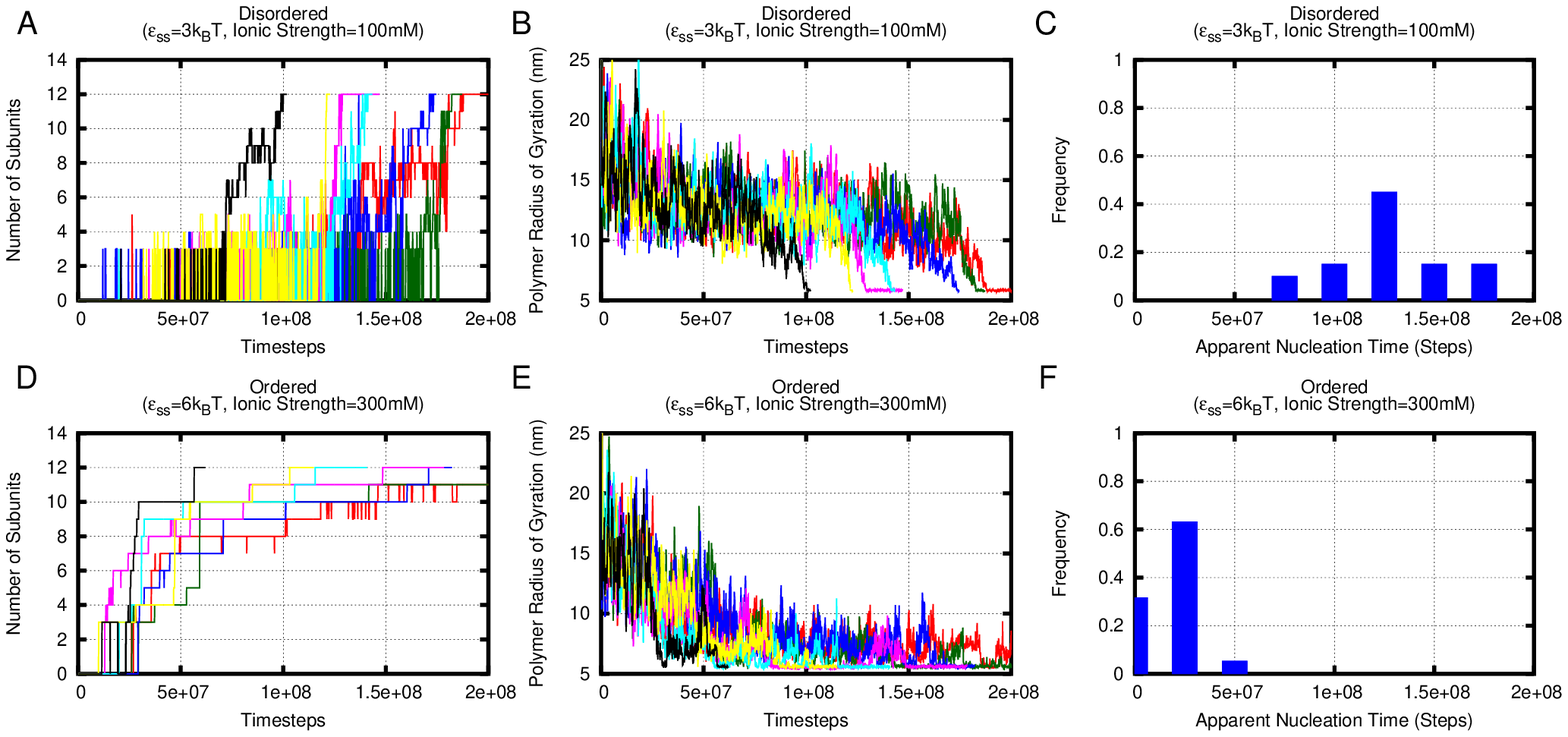}}
\caption{Examples of individual trajectories showing the number of subunits in the largest cluster for an \textit{en masse} assembly {\bf (A)} and nucleation-growth assembly {\bf (D)}. {\bf (B,E)} Polymer radius of gyration for the same simulations. {\bf (C,F)} Histograms of apparent nucleation times for these parameters.
\label{examples}
}
\end{figure}

\begin{figure}[p]
\centering{\includegraphics[width=0.95\columnwidth]{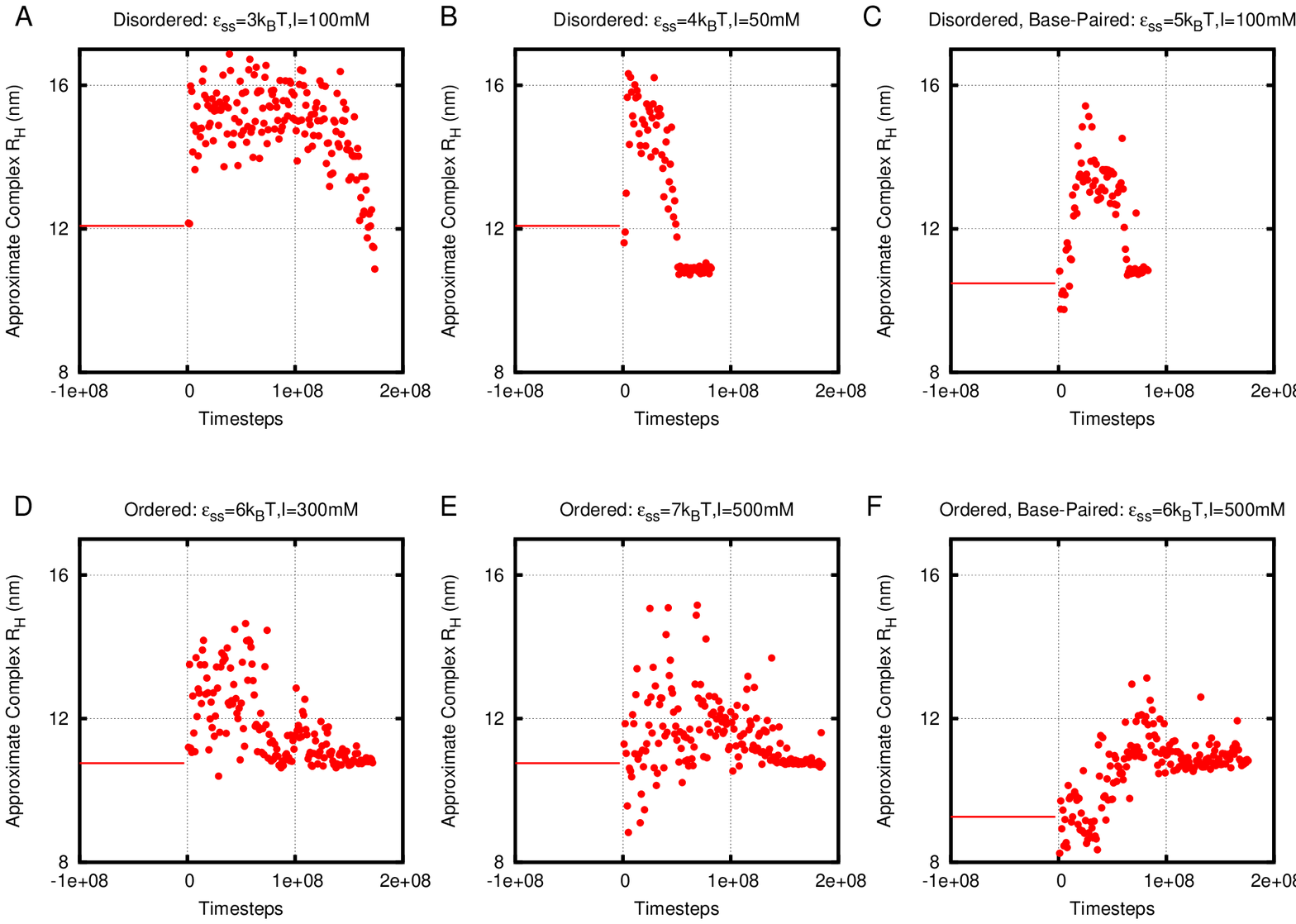}}
\caption{Hydrodynamic radius, $\Rh$, as a function of time for linear polyelectrolytes and the model NA described in Ref. \cite{Perlmutter2013}, which accounts for some effects of base pairing. \textit{En masse} assembly leads to a larger increase in $\Rh$ for assembly around both linear {\bf (A,B)} and base paired {\bf (C)} polymers, compared to nucleation-and-growth trajectories around linear {\bf (D,E)} and base-paired {\bf (F)} polymers.}
\label{radiusSupplement}
\end{figure}

\textit{Subunit-Polymer interaction.} Our method of estimating the subunit-polymer binding free energy is also similar to that presented in our previous work \cite{Elrad2010, Hagan2011}. In these simulations, the subunit-subunit attraction is eliminated (i.e. $\ess{=}0\kt$), and the average number of subunits adsorbed to the polymer is measured as a function of the subunit concentration and salt concentration (Fig.~\ref{affinityDensity}A). With sufficient data, the dissociation constant and free energy as a function of salt concentration could be determined using the McGhee-von Hippel formulation, wherein a bound ligand occupies multiple binding sites on the polymer \cite{Mcghee1974}. In our previous simulations the number of binding sites occupied by a subunit was determined by the model \cite{Elrad2010, Hagan2011}, whereas here the number of polymer sites occupied per subunit emerges from collective interactions. As our estimate, we use the number of subunits per polymer length for the maximal saturation observed, which for our subunits containing 25 ARM charges (5 ARMs of length 5 each) results in $p=22$. This leads to estimates in the free energy of subunit-polymer binding of $\sim3-6\kt$ for $\Csalt = 500 - 100$ mM.  However, we found systematic variations of our estimate for changing total subunit concentration, suggesting that our sampling was incomplete. 

\subsection{Constructing the Markov state model}
\label{sec:msm}
Markov state models (MSMs) have been used extensively to study protein folding \cite{Bowman2011, Bowman2010, Bowman2009, Swope2004, Swope2004a, Prinz2011, Park2006, Pande2010, Noe2009, Lane2011, Jayachandran2006, Chodera2011, Hinrichs2007, Chodera2007, Deuflhard2005}, and are one of the few methods that can describe out-of-equilibrium dynamical processes that include rare events, such as a nucleation barrier crossing. Our strategy for using MSMs to study capsid self-assembly is described in detail in Ref. \cite{Perkett2014}; we summarize the key points here. We construct an MSM by running many short (relative to the assembly timescale) simulations using a ratcheting procedure, which starts simulations from configurations based on the most poorly sampled states. System configurations are partitioned into states such that configurations which interconvert rapidly are collected in the same state while those which interconvert slowly are in separate states. The separation of timescales provided by such decomposition ensures that the system behaves Markovian beyond the relatively short intra-state relaxation timescales.  Inter-state transition probabilities are then measured after a `lag time' which is longer than intra-state relaxation times.
We follow the simplest approach described in Ref. \cite{Perkett2014}, in which states are defined by the largest cluster size $n$ and the total number of subunits adsorbed to the polyelectrolyte $\nads$. Using this state decomposition produced 74 states for the slow nucleation parameters ($\ess=5\kt$, $\Csalt=500$mM) with a lag time of $4 \times 10^5$ simulation steps sufficient to maintain the Markov property.

\begin{figure}[p]
\centering{\includegraphics[width=0.95\columnwidth]{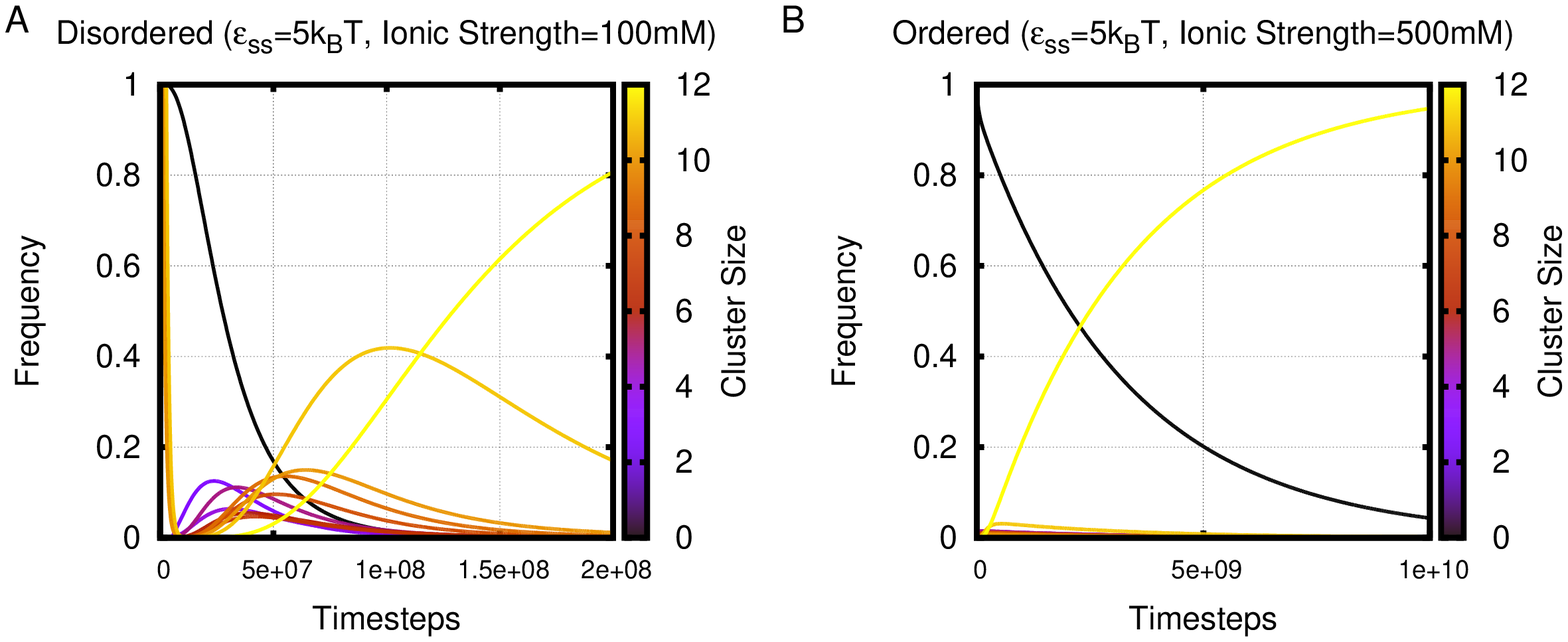}}
\caption{Frequency of structures from Markov State Models (MSMs) of capsid assembly at two parameters. {\bf (A)} Results from an MSM built from unbiased simulations at parameters which result in disordered intermediates. {\bf (B)} Results of MSM built from ratcheting simulations (see section \ref{sec:msm}), at parameters which result in approximately two-state kinetics.}
\label{twoState}
\end{figure}

\textit{MSM Calculations.} The transition probability matrix $\mathbf{T}(\tau)$ is calculated by column-normalizing the count matrix $\mathbf{C}(\tau)$, in which each element $C_{ji}$ gives the total number of transitions from state $i$ to state $j$ measured at a lag time $\tau$. The count matrix is calculated from the many, relatively short, trajectories run in parallel using the ratcheting procedure described below. After diagonalizing $\mathbf{T}(\tau)$, the time-dependent state probabilities can be written as
\begin{align}
\vec{P}(t;\tau) & = \sum_{i=1}^{N} \ket{i} \bra{i} \ket{\vec{P}(0)} e^{-\lambda_i t} \nonumber \\
\lambda_i &= -\log(\omega_i)/\tau
\label{eq:msm}
\end{align}
where $\omega_i$ is the $i^{\text{th}}$ eigenvalue of $\mathbf{T}(\tau)$ and $\bra{i}$ and $\ket{i}$ are the corresponding left/right eigenvectors, which are assumed to be normalized. Since $\mathbf{T}(\tau)$ is generally not Hermitian the left and right eigenvectors are not equivalent. There is only one unit eigenvalue, whose associated right eigenvector corresponds to the equilibrium distribution, while all other eigenvalues are positive and real \cite{Swope2004a}. The average value of an order parameter, $Q$, can then be calculated as a function of time from
\begin{equation}
\bar{Q}(t) = \sum_{i=1}^{N} P_i(t;\tau) Q_i
\label{eq:msm2}
\end{equation}
with $Q_i$ as the order parameter value for state $i$ and $P_i(t;\tau)$ as the time dependent probability of state $i$ (from Eq.~\ref{eq:msm}).  Eq.~\ref{eq:msm2} was used to calculate the mean cluster size $n$ and mean number adsorbed $\nads$ in SI Fig.~\ref{snapsGrowth}C. In order to calculate the probability of a given cluster size as a function of time in SI Fig. \ref{twoState}, the same procedure was followed except that all states for a given $n$ were lumped together to give a single curve for each $n$.

\textit{Ratcheting Procedure.} Because of the Markov property, starting coordinates for simulations can be chosen to efficiently generate good statistics for all of the relevant transition elements. Many simulations are run in parallel for a time $t_\text{s}$ , which must be longer than the lag time $\tau$, but can be much shorter than the longest relaxation timescale. For the slow nucleation calculation in this paper, we continuously ran $100$ simulations in parallel starting from an unassembled bath of subunits.  New simulations were seeded from the states with the fewest number of simulation starts, biasing sampling to unexplored regions of state space.  Our ratcheting procedure initially grouped states only by the number of subunits adsorbed to polymer $\nads$. However, after discovering states for each cluster size on the pathway to assembly, we ran a large number of simulations in parallel seeded from a random selection of existing states. This crude, but effective approach could be improved by using adaptive sampling \cite{Bowman2010}, which explicitly considers the error in the MSM when seeding new simulations. A total of $6.3\times 10^9$ simulation steps were used to build the MSM.

\end{document}